\def\ie{{\it i.e.}}

\def\~{{$\tilde{\phantom{a}}$}}

\documentclass [12pt] {article}
\usepackage{epsfig}
\usepackage{color}

\def\thebibliography#1{\section{References}\markboth
 {REFERENCES}{REFERENCES}\list
 {[\arabic{enumi}]}{\settowidth\labelwidth{[#1]}\leftmargin\labelwidth
 \advance\leftmargin\labelsep
 \usecounter{enumi}}
 \def\newblock{\hskip .11em plus .33em minus -.07em}
 \sloppy
 \sfcode`\.=1000\relax}
\def\upcite#1{\raise6pt\hbox{\scriptsize
\cite{#1}}}
\def\lsim{\mathrel {\vcenter {\baselineskip 0pt \kern 0pt
    \hbox{$<$} \kern 0pt \hbox{$\sim$} }}}
\def\gsim{\mathrel {\vcenter {\baselineskip 0pt \kern 0pt
    \hbox{$>$} \kern 0pt \hbox{$\sim$} }}}
\def\gtlt{\mathrel {\vcenter {\baselineskip 0pt \kern 0pt
    \hbox{$>$} \kern 0pt \hbox{$<$} }}}

\def\hline{\noalign{\hrule \vskip2pt}}

\def\|{\ifmmode\Vert\else \char`\|\fi}
\ifx\oldzeta\undefined                          % hasn't been done yet, so 
  \let\oldzeta=\zeta                            % save old definiton
  \def\zzeta{{\raise 2pt\hbox{$\oldzeta$}}}     % make new definition
  \let\zeta=\zzeta                              % and attatch it
\fi

\ifx\oldchi\undefined                           % hasn't been done yet, so 
  \let\oldchi=\chi                              % save old definiton
  \def\cchi{{\raise 2pt\hbox{$\oldchi$}}}       % make new definition
  \let\chi=\cchi                                % and attatch it
\fi

\def\frac#1#2{{#1 \over #2}}

\def\half{\ifinner {\scriptstyle {1 \over 2}}
   \else {1 \over 2} \fi}

\def\simge{\mathrel{%
   \rlap{\raise 0.511ex \hbox{$>$}}{\lower 0.511ex \hbox{$\sim$}}}}
\def\simle{\mathrel{
   \rlap{\raise 0.511ex \hbox{$<$}}{\lower 0.511ex \hbox{$\sim$}}}}

\def\buildchar#1#2#3{{\null\!                   % \null, cancel space
   \mathop#1\limits^{#2}_{#3}                   % #1, #2 above, #3 below
   \!\null}}                                    % cancel space, \null
\def\overcirc#1{\buildchar{#1}{\circ}{}}

\def\slashchar#1{\setbox0=\hbox{$#1$}           % set a box for #1 
   \dimen0=\wd0                                 % and get its size
   \setbox1=\hbox{/} \dimen1=\wd1               % get size of /
   \ifdim\dimen0>\dimen1                        % #1 is bigger
      \rlap{\hbox to \dimen0{\hfil/\hfil}}      % so center / in box
      #1                                        % and print #1
   \else                                        % / is bigger
      \rlap{\hbox to \dimen1{\hfil$#1$\hfil}}   % so center #1
      /                                         % and print /
   \fi}                                         %

\def\subrightarrow#1{%                          % #1 under arrow
  \setbox0=\hbox{%                              % set a box
    $\displaystyle\mathop{}%                    % no mathop
    \limits_{#1}$}%                             % just limits
  \dimen0=\wd0%                                 % get width
  \advance \dimen0 by .5em%                     % add a bit
  \mathrel{%                                    % space like =
    \mathop{\hbox to \dimen0{\rightarrowfill}}% % arrow to width
       \limits_{#1}}}                           % text below

\def\overlay#1#2{\ifmmode%
\setbox0=\hbox{$#1$}%
\setbox1=\hbox to\wd0{\hss$#2$\hss}\else%
\setbox0=\hbox{#1}%
\setbox1=\hbox to\wd0{\hss#2\hss}\fi%
#1\hskip-\wd0\box1 }

\def\pmb#1{\leavevmode\setbox0=\hbox{#1}%
\kern-.02em\copy0\kern-\wd0
\kern.04em\copy0\kern-\wd0
\kern-.02em\raise.04em\box0 }

\def\vereq#1#2{\lower3pt\vbox{\baselineskip1.5pt \lineskip1.5pt
\ialign{$\m@th#1\hfill##\hfil$\crcr#2\crcr\sim\crcr}}}

\def\tensor#1{\protect\@ontopof{#1}{\leftrightarrow}{1.15}\mathord{\box2}}
\def\overstar#1{\protect\@ontopof{#1}{\ast}{1.15}\mathord{\box2}}
\def\overdots#1{\protect\@ontopof{#1}{\cdots}{1.0}\mathord{\box2}}
\def\overcirc#1{\protect\@ontopof{#1}{\circ}{1.2}\mathord{\box2}}
\def\loarrow#1{\protect\@ontopof{#1}{\leftarrow}{1.15}\mathord{\box2}}
\def\roarrow#1{\protect\@ontopof{#1}{\rightarrow}{1.15}\mathord{\box2}}

\def\@ontopof#1#2#3{%
{\mathchoice
{\@@ontopof{#1}{#2}{#3}\displaystyle\scriptstyle}%
{\@@ontopof{#1}{#2}{#3}\textstyle\scriptstyle}%
{\@@ontopof{#1}{#2}{#3}\scriptstyle\scriptscriptstyle}%
{\@@ontopof{#1}{#2}{#3}\scriptscriptstyle\scriptscriptstyle}%
}%
}

\def\@@ontopof#1#2#3#4#5{%
\setbox0=\hbox{$#4#1$}%
\setbox1=\hbox{$#5#2$}%
\setbox2=\hbox{}\ht2=\ht0 \dp2=\dp0 %
\ifdim\wd0>\wd1 %
\setbox1=\hbox to\wd0{\hss\box1\hss}%
\mathord{\rlap{\raise#3\ht0\box1}\box0}%
\else   %
\setbox1=\hbox to.9\wd1{\hss\box1\hss}%
\setbox0=\hbox to\wd1{\hss$#4\relax#1$\hss}%
\mathord{\rlap{\copy0}\raise#3\ht0\box1}%
\fi
}%

\def\lambdabar{\protect\@lambdabar}
\def\@lambdabar{%
\relax
\bgroup
\def\@tempa{\hbox{\raise.73\ht0
\hbox to0pt{\kern.25\wd0\vrule width.5\wd0
height.1pt depth.1pt\hss}\box0}}%
\mathchoice{\setbox0\hbox{$\displaystyle\lambda$}\@tempa}%
{\setbox0\hbox{$\textstyle\lambda$}\@tempa}%
{\setbox0\hbox{$\scriptstyle\lambda$}\@tempa}%
{\setbox0\hbox{$\scriptscriptstyle\lambda$}\@tempa}%
\egroup
}

\def\corresponds{{\lower.2ex\hbox{=}}{\rm\kern-.75em^\triangle}}
\def\succsim{\succ\kern-.9em_\sim\kern.3em}
\def\precsim{\prec\kern-1em_\sim\kern.3em}
\def\slantfrac#1#2{\kern1em^{#1}\kern-.3em/\kern-.1em_{#2}}

\begin{document}
                                                                
\begin{center}
{\Large\bf Maximal Gravity at the Surface of an Asteroid}
\\
\medskip

Kirk T.~McDonald
\\
{\sl Joseph Henry Laboratories, Princeton University, Princeton, NJ 08544}
\\
(February 18, 2002)
\end{center}

\section{Problem}

What is the shape of an asteroid of given uniform density and given total
mass such that the
force of gravity is maximal for one point on its surface?  Compare the maximal
gravity with that of a sphere of the same mass and density.

\section{Solution}

This problem is from Yakov Kantor's Physics Quiz site,
http://star.tau.ac.il/QUIZ/

If the asteroid is spherical with radius $a$ and mass $M$, the force of gravity on
a test mass $m$ is
\begin{equation}
F = {G M m \over a^2}\, ,
\label{s1}
\end{equation}
everywhere on the surface, where $G$ is Newton's constant of gravitation.

Can some other shape of the asteroid result in a larger force?  
Let the test mass be at the origin, which is on the surface of the asteroid, and
define the $z$ axis to be along the direction of the desired maximal force on the
test mass.  It is ``obvious'' that this $z$ axis is an axis of symmetry of the
asteroid: if the asteroid were not symmetric about the $z$ axis, the $z$ component
of the vector force $G m m' \hat{\bf r} / r^2$ would be increased by moving material
from larger to smaller transverse distances from the axis, until the asteroid is
axially symmetric.

We seek the functional form $x = x(z)$, where $0 < z < z_0$, that generates a
surface of revolution about the $z$ axis, the surface of the asteroid, 
that maximizes the (axial) force on test mass $m$ at the origin.  The axial force due
to a ring of extent $dx\ dz$ that passes through point $(x,z)$ is
\begin{equation}
G m \rho\ 2 \pi x\ dx\ dz {1 \over x^2 + z^2} {z \over \sqrt{x^2 + z^2}}\, ,
\label{s4a}
\end{equation}
so the total force is 
\begin{equation}
F = 2 \pi G m \rho \int_0^{z_0} dz \int_0^{x(z)} x\ dx {z \over (x^2 + z^2)^{3/2}}
 = {3 G M m \over 2 a^3} \int_0^{z_0} dz \left( 1 -{z \over (x^2(z) + z^2)^{1/2}}
\right) ,
\label{s2}
\end{equation}
where we suppose that the asteroid has the same
density as a sphere of radius $a$ and mass
$M$, \ie, $\rho = 3 M / 4 \pi a^3$.  Hence, the volume of the asteroid is
constrained to have value
\begin{equation}
V = \pi \int_0^{z_0} dz\ x^2(z) = {4 \pi a^3 \over 3}\, .
\label{s3}
\end{equation}

This suggests use of the calculus of variations, although there is the
complication that the value of the endpoint $z_0$ is unknown. 

In principle, the shape $x(z)$ might involve nonzero values at $z = 0$ and $z_0$,
\ie, planar surfaces that bound the asteroid.  However, we readily convince ourselves
that such a geometry could not maximize the $z$ component of the gravitational
force; material in the plane $z = 0$ contributes nothing to the $z$ component of the
force, while material at $x > 0$ in the plane $z = z_0$ would be more useful if it
were moved to a smaller $z$.  Thus, we understand that $x(0) = 0 = x(z_0)$ for the
desired shape function.

\subsection{An Intuitive Solution}

However, we first consider the related question: what is
the shape of a thin shell of matter such that all points on the shell contribute
the same axial force on our test mass at the origin?  For a shell described by
$x(z)$, the answer is
already contained in the form of integral (\ref{s2}).  Namely, we desire that
\begin{equation}
{z \over (x^2(z) + z^2)^{3/2}} = {\rm const.} = {1 \over z_0^2}\, ,
\label{s4}
\end{equation}
where we have evaluated the constant at the intercept $z_0$ of the shell with the
$z$ axis.  On rearranging eq.~(\ref{s4}), we have
\begin{equation}
x^2(z) = z^{2/3} z_0^{4/3} - z^2.
\label{s5}
\end{equation}
It is noteworthy that this form also intersects the $z$ axis at the origin, so is
a satisfactory candidate for the shape of the asteroid (whose surface surely must
touch the test mass at the origin).  Inserting this form in
eq.~(\ref{s3}) for the volume, we find that $z_0 = \sqrt[3]{5} a = 1.7 a$.  Using 
eq.~(\ref{s5}) in eq.~(\ref{s2}) for the axial force, we find
\begin{equation}
F = {3 G M m z_0 \over 5 a^3} = {3 \sqrt[3]{5} G M m \over 5 a^2} 
= 1.026 {G M m \over a^2}\, . 
\label{s6}
\end{equation}

This argument does not prove that eq.~(\ref{s5}) describes the shape of the 
maximal gravity asteroid, but it does show that the shape is not a sphere.
The simplicity of the argument suggests that the result is
in fact correct, but a proof of this is still desired. 

\subsection{Solution via the Calculus of Variations}

We wish to find the form $x(z)$ that maximizes integral $F$ of eq.~(\ref{s2})
subject to the constraint that integral $V$ of eq.~(\ref{s3}) has the stated value,
while permitting the endpoint $z_0$ to vary.  To proceed via the calculus of
variations, we consider the integral
\begin{equation}
I = \int_0^{z_0} f(z,x(z))\ dz,
\label{s7}
\end{equation}
where
\begin{equation}
f = 1 -{z \over (x^2 + z^2)^{1/2}} - \lambda x^2,
\label{s8}
\end{equation}
which combines the integrands of eqs.~(\ref{s2}) and (\ref{s3}) using the Lagrange
multiplier $\lambda$.

For a fixed endpoint $z_0$, integral $I$ is at an extremum provided $f$ obeys the
Euler-Lagrange equations
\begin{equation}
{\partial f \over \partial x} = {d \over dz} {\partial f \over \partial x'}\, .
\label{s9}
\end{equation}
Since $f$ does not depend on $x' = dx/dz$ in this problem, we simply have
\begin{equation}
0 = {\partial f \over \partial x} = {x z \over (x^2 + z^2)^{3/2}} - 2 \lambda x,
\label{s10}
\end{equation}
or
\begin{equation}
{z \over (x^2 + z^2)^{3/2}} = 2 \lambda.
\label{s11}
\end{equation}
As in eq.~(\ref{s4}), the constant $2 \lambda$ is clearly $1 / z_0^2$, since $x(z_0)
= 0$.  Again, we obtain eq.~(\ref{s5}) as the shape of the asteroid.  And again, the
constant $z_0$ (\ie, the multiplier $\lambda$) must be set to $\sqrt[3]{5} a$
to satisfy the volume constraint (\ref{s3}), which completes the solution.

The shape (\ref{s5}) of the maximal gravity asteroid is shown below.  It is 
noteworthy that there is no cusp at either $z = 0$ or $z_0$.

\vspace{0.1in}
\centerline{\includegraphics[width=4in]{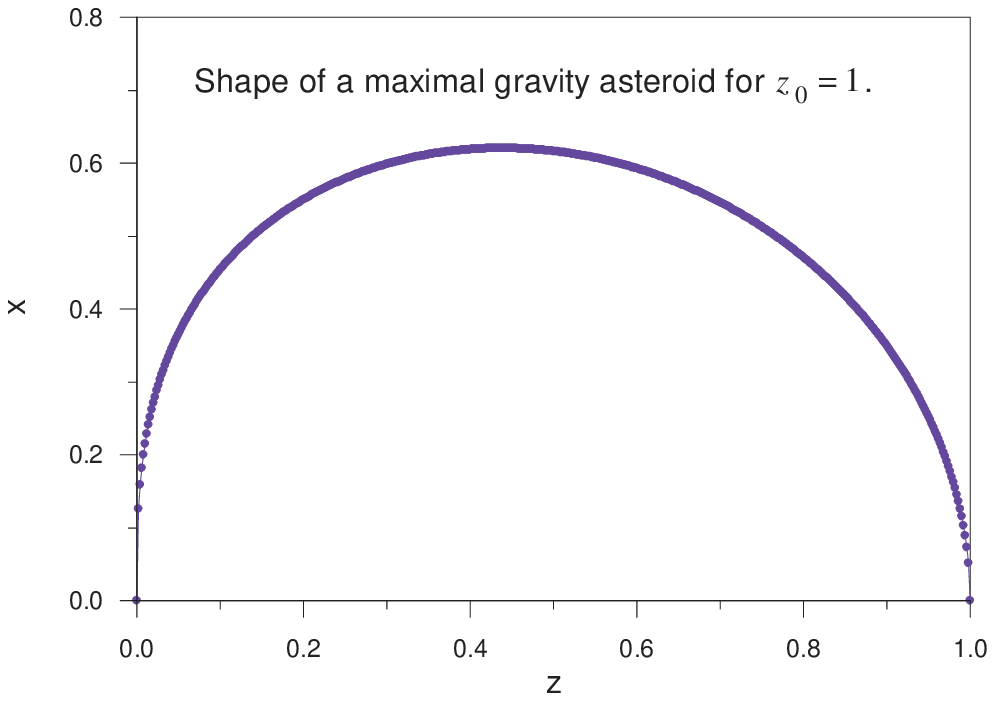}}

\subsection{A Mathematical Footnote}

We have not directly included in our calculus of variations the fact that the endpoint $z_0$ is free to move along the $z$ axis, \ie, along the curve $g(z) = 0$.
In, for example, sec.~4.3 of \cite{Fox}
we read that the freedom to vary the endpoint in the maximization of
integral $I$ leads to the additional relation
\begin{equation}
f + (g' - x'){\partial f \over \partial x'} = 0,
\label{s12}
\end{equation}
at the variable endpoint $z_0$,
where $g(z)$ describes the curve on which the endpoint $z_0$ can vary.  
Applying this to the present problem, we must have $f(z_0) = 0$, which requirement
is satisfied by the condition $x(z_0) = 0$ that has already been included in the
analysis of sec.~2.2.  However, it is pleasant to have a formal confirmation of
our earlier ``intuitive'' justification of this condition.

\subsection{Further Comment on the Shape the Asteroid}

The form (\ref{s5}) for the shape of a maximal gravity asteroid differs only slightly
from that of a sphere, as shown in the figure on p.~3.  Starting from a sphere,
the force of gravity at, say, the ``north'' pole is increased by moving material 
from the ``southern'' hemisphere into the ``northern'' hemisphere.

We verify that this is the right thing to do by a simple calculation.  Namely, we
compare the $z$ component of the force of gravity at the 
north pole due to a point mass on the surface of a sphere of radius $a$ at polar angle $\theta$ with that due to a
point mass at angle $\pi - \theta$.  The claim is that the point at $\theta$ 
contributes more to the force than the point at $\pi - \theta$, so that it would be
favorable to move some of the mass to $\theta < \pi / 2$.

The distance from the pole to the point at angle $\theta$ is $r = 2 a \sin\theta / 2$.
The $z$ component of the force of gravity due to a point mass at angle $\theta$ is proportional to 
\begin{equation}
F_z(\theta) \propto {\Delta z \over r^3} 
= {a (1 - \cos\theta) \over 8 a^3 \sin^3 \theta / 2}
= {1 \over 4 a^2 \sin \theta / 2}\, ,
\label{13}
\end{equation}
so that the force due to a point at angle $\pi - \theta$ is related by
\begin{equation}
F_z(\pi - \theta) \propto {1 \over 4 a^2 \cos \theta / 2}\, .
\label{14}
\end{equation}
The ratio,
\begin{equation}
{F_z(\pi - \theta) \over F_z(\theta)} = \tan \theta / 2,
\label{15}
\end{equation}
is less than one for all $\theta < \pi / 2$.  Thus, the force of gravity at the
north pole would be
increased by moving mass from angle $\pi - \theta$ to angle $\theta$ on the surface
of the asteroid, as claimed.

While this argument gives a qualitative feel as to how a spherical asteroid should
be reshaped to become a maximal-gravity asteroid, it does not provide a
quantitative prescription of how much mass should be moved.  For that, the
arguments of secs.~2.1 and 2.2 are better.


\begin{thebibliography}{99}

\bibitem{Fox}
C.~Fox,
{\em An Introduction to the Calculus of Variations}
(Oxford U.P., Oxford, 1950; Dover Publications, New York, 1987).

\end{thebibliography}
\end{document}